\documentclass[aps,pra,twocolumn,superscriptaddress,showpacs,floatfix]{revtex4}
\usepackage{epstopdf}
\usepackage{graphicx}
\usepackage{dcolumn}
\usepackage{bm}
\usepackage{amsfonts}
\usepackage{amsmath}

\begin{document}
\renewcommand{\thefigure}{\arabic{figure}}
\def\be{\begin{equation}}
\def\ee{\end{equation}}
\def\ber{\begin{eqnarray}}
\def\eer{\end{eqnarray}}

\title{Ground-state and dynamical properties of two-dimensional dipolar Fermi liquids}
\date{\today}
\author{Saeed H. Abedinpour}
\email{abedinpour@iasbs.ac.ir}
\affiliation{Department of Physics, Institute for Advanced Studies in Basic Sciences (IASBS), Zanjan 45137-66731, Iran}
\author{Reza Asgari}
\affiliation{School of Physics, Institute for Research in Fundamental Sciences (IPM), Tehran 19395-5531, Iran}
\author{B. Tanatar}
\affiliation{Department of Physics, Bilkent University, Bilkent, 06800 Ankara, Turkey}
\author{Marco Polini}
\affiliation{NEST, Istituto Nanoscienze-CNR and Scuola Normale Superiore, I-56126 Pisa, Italy}

\begin{abstract}
We study the ground-state properties of a two-dimensional spin-polarized fluid of dipolar fermions within the
Euler-Lagrange Fermi-hypernetted-chain approximation. Our method is
based on the solution of a scattering Schr\"odinger equation for the
``pair amplitude" $\sqrt{g(r)}$, where $g(r)$ is the pair distribution function. A key ingredient in our theory is the effective pair potential,
which includes a bosonic term from Jastrow-Feenberg correlations
and a fermionic contribution from kinetic energy and exchange, which is tailored
to reproduce the Hartree-Fock limit at weak coupling. Very good
agreement with recent results based on quantum Monte Carlo simulations
is achieved over a wide range of coupling constants up to the liquid-to-crystal quantum phase transition (QPT).
Using a certain approximate model for the dynamical
density-density response function, we furthermore demonstrate
that: i) the liquid phase is stable towards the formation of
density waves up to the liquid-to-crystal QPT and ii) an
undamped zero-sound mode exists for any value of the interaction strength, down to infinitesimally weak couplings.
\end{abstract}

\maketitle

\section{Introduction}\label{sect:intro}

Recent experimental breakthroughs in trapping and cooling polar
molecules and atoms with large permanent magnetic moments
has triggered an immense theoretical interest in quantum dipolar
fluids~\cite{ref:baranov_phyrep08,baranov_arxiv_2012,ref:lahaye_rpp09,aikawa_prl_2012,lu_prl_2012}.
Unlike the usual van der Waals interaction between atoms, which can be
replaced by a contact Fermi pseudo-potential at ultra-low
temperatures~\cite{ref:pethick_book}, the dipole-dipole interaction is
long ranged and anisotropic. It is therefore natural to expect more exotic
phases in ultra-cold dipolar gases. While one of the greatest advantages of
short-range interactions is their tunability through Feshbach
resonances~\cite{ref:pethick_book,chin_rmp_2010}, techniques have been
proposed~\cite{giovanazzi_prl_2002} for controlling both strength and
sign of dipolar interactions as well.

As already mentioned, the inter-particle interaction between polarized ({\it i.e.} dipoles aligned in the same direction)
dipoles has two important features: i) it is long-ranged, {\it i.e.} it decays like $1/r^3$ at large
distances, and ii) it is anisotropic. In particular, it is
repulsive for dipoles aligned side-by-side and
is attractive for dipoles aligned head-to-toe.

It is worth mentioning that the attractive part of dipole-dipole interactions can
drive dipolar fluids towards instabilities. In alkali-metal
diatomic molecules such as K-Rb, Li-Na, etc., some chemically reactive channels,
which are energetically favorable, exist and lead to particle recombination and two-body losses in the
gas~\cite{zuchowski_pra_2010, baranov_arxiv_2012}.

A very simple method for stabilizing dipolar gases is to confine
them into low-dimensional geometries. For example, a trap with
pancake geometry together with a polarizing field, which aligns the
dipoles along the direction of transverse confinement, simulates a stable two-dimensional
(2D) system with purely repulsive and isotropic dipolar interactions of the form
\be\label{eq:udd_2d}
v_{\rm dd}(r) = \frac{C_{\rm dd}}{4\pi}~\frac{1}{r^3}~.
\ee
Here $C_{\rm dd}$ is the dipole-dipole coupling constant, which
depends on the microscopic origin of the interaction:
{\it e.g.}, it is $d^2/\epsilon_0$ for particles with permanent
electric dipole $d$ and $\mu_0 M^2$ for particles with permanent
magnetic dipole $M$ (here $\epsilon_0$ and $\mu_0$ are the permittivity
and permeability of vacuum, respectively).

Ground-state properties and collective modes of 2D dipolar fermions
have been addressed in a number of studies~\cite{bruun_prl_2008, chan_pra_2010, lu_pra_2012, matveeva_arxiv_2012, li_prb_2010, sieberer_pra_2011, babadi_pra_2011, parish_prl_2012, marchetti_arxiv_2012, yamaguchi_pra_2010, sun_prb_2010,  babadi_prb_2011, zinner_epjd_2011}. For their particular relevance to this Article we highlight the following two recent studies~\cite{lu_pra_2012,matveeva_arxiv_2012} of a 2D dipolar Fermi gas (DFG) with isotropic interactions as in Eq.~(\ref{eq:udd_2d}). Lu and Shlyapnikov~\cite{lu_pra_2012} have calculated a number of Fermi-liquid properties of a weakly interacting 2D DFG. In particular, these authors have presented several exact results up to second order in a natural dimensionless coupling constant, which we have introduced below in  Eq.~(\ref{eq:couplingconstant}). More recently, Matveeva and Giorgini~\cite{matveeva_arxiv_2012} have carried out quantum Monte Carlo (QMC) simulations of a 2D DFG, presenting in particular results for the phase diagram of this system over a wide range of coupling constants. These studies pose severe bounds on any microscopic theory of 2D DFGs.

In this Article we present a theoretical study of ground-state and dynamical
properties of a 2D DFG with average
density $n$. Our main focus is on the pair distribution function (PDF) $g(r)$, which is often referred to as ``Pauli-Coulomb hole".
This is defined~\cite{Pines_and_Nozieres,Giuliani_and_Vignale} so that the quantity $2\pi n g(r)dr$
gives the average number of dipoles lying within a circular shell
of radius $dr$ centered on a ``reference'' dipole sitting at the origin.
We present a self-consistent semi-analytic theory of the PDF,
which incorporates many-body exchange and correlation effects, thereby allowing us to
explore the physics of the system at strong coupling. Our approach,
which is based on the so-called Euler-Lagrange Fermi-hypernetted-chain (FHNC)
approximation at zero temperature~\cite{lantto,zab,report,recentreview}, involves the
solution of a zero-energy scattering Schr\"{o}dinger equation with
a suitable effective
potential~\cite{ref:kallio, davoudi_prb_2003_fermions, davoudi_prb_2003_bosons, asgari_ssc_2004, abedinpour_ssc_2007, abedinpour_dipolar_boson}. This contains a ``bosonic term" from Jastrow-Feenberg correlations and a
``fermionic term" from kinetic energy and exchange, which is tailored to
reproduce the Hartree-Fock (HF) limit at weak coupling and guarantees
the antisymmetry of the fermionic wave function. Furthermore, we use the fluctuation-dissipation theorem~\cite{Pines_and_Nozieres,Giuliani_and_Vignale} and the PDF obtained from the FHNC approximation to calculate the dynamical density-density linear-response function. With this quantity at our disposal, we investigate the possibility of instabilities towards inhomogenous ground states ({\it i.e.} density waves) at strong coupling and the existence of  a ``zero sound" mode~\cite{Pines_and_Nozieres} in a 2D DFG. Our results are severely benchmarked against the findings of Refs.~\cite{lu_pra_2012,matveeva_arxiv_2012}.

This Article is organized as follows. In Sect.~\ref{sect:theory}
we present our model and the self-consistent method we use to calculate in an accurate manner the PDF of a 2D DFG.
In Sect.~\ref{sect:CDW} we discuss a number of approximations we make to derive the dynamical density-density linear-response function of a 2D DFG and explain how this can be used to examine the tendency towards a density-wave instability and the emergence of a collective zero-sound mode due to many-body effects. Sect.~\ref{sect:results} collects our main numerical results, while Sect.~\ref{sect:concl} contains a brief summary of our main findings.

\section{Scattering theory for the Pauli-Coulomb hole}
\label{sect:theory}
We consider a spin-polarized 2D DFG described by the following first-quantized Hamiltonian~\cite{astrakharchik_prl_2007}:
\begin{equation}\label{eq:hamil}
{\cal H} =-\frac{\hbar^2}{2m}\sum_i\nabla^{2}_{{\bm r}_i} + \sum_{i<j}v_{\rm dd}(|{\bm r}_{i} - {\bm r}_{j}|)~,
\end{equation}
where $m$ is the mass of a dipole and the bare dipole-dipole interaction has been introduced above in Eq.~(\ref{eq:udd_2d}).
The ground-state properties of the Hamiltonian~(\ref{eq:hamil}) are governed by a single
dimensionless parameter:
\be\label{eq:couplingconstant}
\lambda=k_{\rm F} r_0~,
\ee
where $r_0=m C_{\rm dd}/(4\pi\hbar^2)$ is a characteristic length scale
and $k_{\rm F}=\sqrt{4\pi n}$ is the Fermi wave number, $n$ being
the 2D average density.

In order to calculate the ground-state properties of the Hamiltonian
(\ref{eq:hamil}), we use the FHNC~\cite{lantto,zab,report} approximation
at zero temperature. In what follows we first present our theory at the
simplest level (which works well in the perturbative regime $\lambda \ll 1$) and
then transcend it to obtain accurate results at strong coupling
($\lambda \gg 1$).

With the zero of energy taken at the chemical potential, one can write a formally exact differential equation for the PDF~\cite{davoudi_prb_2003_fermions,davoudi_prb_2003_bosons}:
\begin{equation}\label{eq:scat}
\left[-\frac{\hbar^2}{m}\nabla^2_{\bm r} + V_{\rm eff}(r)\right]\sqrt{g(r)}=0~.
\end{equation}
We write the effective scattering potential $V_{\rm eff}(r)$ as the sum of three contributions:
\be\label{eq:effectivepot}
V_{\rm eff}(r)=v_{\rm dd}(r)+W_{\rm B}(r)+W_{\rm F}(r)~.
\ee
Here $v_{\rm dd}(r)$ is the bare repulsive dipole-dipole
interaction in Eq.~(\ref{eq:udd_2d}) while the bosonic contribution to the scattering potential, $W_{\rm B}(r)$, is defined, at the
level of the so called ``FHNC/0'' approximation, by the following
equation~\cite{ref:chakraborty}:
\be\label{eq:wb}
\left.W_{\rm B}(k)\right|_{{\rm FHNC}/0} \equiv -\frac{\epsilon(k)}{2 n}\left[2S(k)+1\right]\left[\frac{S(k)-1}{S(k)}\right]^2~.
\ee
In writing Eq.~(\ref{eq:wb}) we have introduced the Fourier transform
(FT) $W_{\rm B}(k)$ of $W_{\rm B}(r)$ according to
\be
{\rm FT}[F(r)] \equiv \int d^2{\bm r}~F(r) \exp{(i{\bm k}\cdot {\bm r})}~.
\ee
Furthermore, $\epsilon(k)=\hbar^2k^2/(2m)$ is the single-particle
energy and $S(k)$ is the instantaneous or ``static'' structure
factor~\cite{Giuliani_and_Vignale}, $S(k)=1+ n~{\rm FT}[g(r)-1]$.

When $\lambda \sim 1$ the simplest approximation for
$W_{\rm B}(r)$ in Eq.~(\ref{eq:wb}) is inadequate.
Improvements on Eq.~(\ref{eq:wb}) can be sought in two
directions~\cite{ref:apaja_w3}. The FHNC/0 may be transcended by the
inclusion of (i) low-order ``elementary'' diagrams and (ii)
three-body Jastrow-Feenberg correlations.

The contribution from three-body correlations to the bosonic potential is given by~\cite{ref:apaja_w3}:
\ber\label{eq:w3}
W_{\rm B}^{(3)}(k) &=& \frac{1}{4n^2}\int \frac{d^2{\bm q}}{(2\pi)^2}~S(p)S(q)u_{3}({\bm q}, {\bm p},{\bm k})\big\{\nu_{3}({\bm q}, {\bm p}, {\bm k}) \nonumber\\
&+& [E(p)+E(q)]u_{3}({\bm q}, {\bm p}, {\bm k})\big\}~.
\eer
In the previous equation, ${\bm p} = -({\bm q} + {\bm k})$, $E(k)=\epsilon(k)/S(k)$ is the Bijl-Feynman excitation spectrum~\cite{Giuliani_and_Vignale},
\be\label{eq:nu3}
\nu_{3}({\bm q}, {\bm p}, {\bm k}) = (\hbar^2/m)[{\bm k}\cdot{\bm p}\chi(p) + {\bm k}\cdot{\bm q}\chi(q)+{\bm p}\cdot{\bm q}\chi(q)]~,
\ee
and
\ber\label{eq:u3}
u_{3}({\bm q}, {\bm p}, {\bm k}) &=& -\frac{(\hbar^2/2m)}{E(k)+E(p)+E(q)}\nonumber\\
&\times& [{\bm k}\cdot{\bm p}\chi(k)\chi(p) +{\bm p}\cdot{\bm q}\chi(p)\chi(q) \nonumber \\
&+ & {\bm k}\cdot{\bm q}\chi(k)\chi(q)]~.
\eer
In Eqs.~(\ref{eq:nu3})-(\ref{eq:u3}) $\chi(k)=1-1/S(k)$.
We have taken into account higher-order terms that are missed by the
FHNC/0 approximation by assuming that they lead to corrections to
the scattering potential $V_{\rm eff}(r)$. Using the theory developed
by Apaja {\it et al.}~\cite{ref:apaja_w3}, we have supplemented
$\left.W_{\rm B}(k)\right|_{{\rm FHNC}/0}$ in Eq.~(\ref{eq:wb}) by the inclusion of the
three-body potential $W^{(3)}_{\rm B}(k)$:
\be\label{eq:correlation_w3}
\left. W_{\rm B}(k)\right|_{{\rm FHNC}/\alpha3} \equiv \left.W_{\rm B}(k)\right|_{{\rm FHNC}/0} + \alpha(\lambda)W_{\rm B}^{(3)}(k)~.
\ee
If $ \alpha(\lambda)$ is set to unity, the r.h.s. of
Eq.~(\ref{eq:correlation_w3}) defines the so-called ``FHNC/$3$''
approximation. It has been
shown~\cite{asgari_ssc_2004, abedinpour_ssc_2007,abedinpour_dipolar_boson}
that higher-order corrections beyond FHNC/$3$ can be effectively taken
into account by introducing a weighting function $\alpha(\lambda)> 1$.
This approximation has been termed~\cite{abedinpour_dipolar_boson} ``FHNC/$\alpha3$".
A convenient analytical parametrization of the function $\alpha(\lambda)$ for 2D dipolar fluids
can be found in Ref.~\cite{abedinpour_dipolar_boson}. Using the notation of this Article, it reads as follows:
\be\label{eq:parametrization}
\alpha(\lambda) = 1.88+3.26\exp{(-0.046~\lambda^{1.16})}~.
\ee
The previous equation is valid all the way up to the critical coupling ($\lambda \sim 25$) for the liquid-to-crystal quantum phase transition~\cite{matveeva_arxiv_2012}.

We finally turn to describe the last term in Eq.~(\ref{eq:effectivepot}), which is supposed to take care of the fermionic statistics of the problem.
According to the original version of the FHNC theory~\cite{lantto,zab,report}, the ``Fermi potential" $W_{\rm F}(r)$ has a very complicated form. Here we have decided to use a simple but effective recipe, which was first proposed by Kallio
and Piilo~\cite{ref:kallio} for the 3D electron liquid. In this approximate scheme $W_{\rm F}(r)$ is given by the following expression:
\be\label{eq:wf}
W_{\rm F}(r)=\frac{\hbar^2}{m}\frac{\nabla^2_{\bm r}\sqrt{g_{\rm HF}(r)}}{\sqrt{g_{\rm HF}(r)}}-\lim_{\lambda\to 0} W_{\rm B}(r)~,
\ee
where $g_{\rm HF}(r)$ is the well-known~\cite{Pines_and_Nozieres,Giuliani_and_Vignale} 2D HF PDF and $W_{\rm B}(r)$ is the bosonic potential
defined above in Eq.~(\ref{eq:wb}) (at the FHNC/$0$ level) or in Eq.~(\ref{eq:correlation_w3}) (at the FHNC/$\alpha3$ level).
The simple choice in Eq.~(\ref{eq:wf}) guarantees that
the HF limit is recovered exactly in weak coupling $\lambda \to 0$ limit.
The Fermi potential (\ref{eq:wf}) has been extensively investigated for 3D~\cite{davoudi_prb_2003_fermions} and 2D~\cite{asgari_ssc_2004} electron liquids yielding results in excellent agreement with QMC simulation data.

Equations~(\ref{eq:scat})-(\ref{eq:wb}) and~(\ref{eq:wf})
form a closed set of equations, which can be solved numerically in a self-consistent manner
to the desired degree of accuracy. Practical recipes on how to solve this system of equations are discussed
in detail in Ref.~\cite{davoudi_prb_2003_fermions}.

Once the PDF has been calculated, the ground-state
energy per particle of the system, $\varepsilon_{\rm GS}$, can be easily extracted by using the integration-over-the-coupling-constant algorithm~\cite{Giuliani_and_Vignale}:
\be\label{eq:e_gs}
\varepsilon_{\rm GS}=\varepsilon_{0}+\frac{n}{2}\int_{0}^{1}d\gamma\int d^2{\bm r}~v_{\rm dd}(r) g_{\gamma}(r)~,
\ee
where
$\varepsilon_{0}=\varepsilon_{\rm F}/2 = \hbar^2 k^2_{\rm F}/(4m) = \hbar^2 \lambda^2/(4 m r_0^2)$
is the ground-state energy of the non-interacting system, $\varepsilon_{\rm F}$ being the Fermi energy,
and $g_{\gamma}(r)$ is the PDF of an auxiliary system with scaled
dipole-dipole interactions of the form
$v^{(\gamma)}_{\rm dd}(r) = \gamma v_{\rm dd}(r) = \gamma C_{\rm dd}/(4\pi r^3)$.
In practice, the integration over $\gamma$ is carried out by integrating
over the coupling constant $\lambda$.

In Sect.~\ref{sect:results} we present numerical results obtained only within our most elaborate approximation, {\it i.e.} the  FHNC/$\alpha3$ approximation. Nevertheless, for the sake of simplicity, all our numerical results for $g(r)$, $S(k)$, and $\varepsilon_{\rm GS}$ will be labeled by the acronym ``FHNC" (rather than ``FHNC/$\alpha3$").

\section{Linear-response theory, density-wave instabilities, and collective modes}
\label{sect:CDW}

The density-density linear-response function of a many-particle system can be generically written as follows~\cite{Giuliani_and_Vignale}:
\be\label{eq:chinn}
\chi_{nn}(k,\omega)=\frac{\chi_0(k,\omega)}{1- \psi(k,\omega)\chi_0(k,\omega)} \equiv \frac{\chi_0(k,\omega)}{\varepsilon(k,\omega)}~,
\ee
where $\psi(k,\omega)$ is a suitable dynamical effective potential---not to be confused with the FT of the effective potential $V_{\rm eff}(r)$ which enters the zero-energy scattering Schr\"{o}dinger equation (\ref{eq:scat})---and  $\chi_0(k,\omega)$ is the well-known~\cite{Giuliani_and_Vignale,stern_prl_1967} density-density response function of an ideal ({\it i.e.} non-interacting ) 2D Fermi gas.

In the celebrated Random Phase Approximation (RPA)~\cite{Pines_and_Nozieres,Giuliani_and_Vignale}, the effective potential $\psi(k,\omega)$ is brutally approximated with the FT of the bare inter-particle potential, {\it i.e.} $v_{\rm dd}(r)$ in our case. It is very well known~\cite{Pines_and_Nozieres,Giuliani_and_Vignale} that the RPA neglects short-range exchange and correlation effects and that it is intrinsically a weak-coupling theory. It is thus not expected to work well (in reduced spatial dimensions and) for values of the dimensionless coupling constant $\lambda \gtrsim 1$.
One of the main drawbacks of the RPA is that it grossly overestimates the strength of the Pauli-Coulomb hole by predicting large and negative values for $g(r)$ at short distances, thereby violating the fundamental request $g(r)>0$. Moreover, in the context of dipolar Fermi gases, the RPA predicts that the long-wavelength collective excitation spectrum (zero-sound mode) is empathic to the short-range details of the bare interaction potential~\cite{li_prb_2010},
{\it i.e.} the ultraviolet cut-off which is needed to regularize the FT of the bare dipole-dipole potential $v_{\rm dd}(r)$.

In the past sixty years or so, a wide body of literature has been devoted to transcend the RPA, especially in the context of 2D electron liquids in semiconductor heterojunctions~\cite{Giuliani_and_Vignale}. Following the seminal works by Hubbard~\cite{hubbard_1957} and Singwi, Tosi, Land, and Sj\"{o}lander~\cite{STLS} (STLS), one successful route has been based on the use of ``local field factors"~\cite{Giuliani_and_Vignale,LFF} (LFFs). Here we will not use Hubbard or STLS LFFs. (For a successful employment of the STLS approximation in the context of 2D DFGs see, for example, Ref.~\cite{parish_prl_2012}.) In this Article we would like to construct a reliable approximation for the density-density response function $\chi_{nn}(k,\omega)$, which is based on the FHNC theory of the PDF outlined in Sect.~\ref{sect:theory}.

We thus start from the well-known fluctuation-dissipation theorem (FDT)~\cite{Giuliani_and_Vignale}, which relates the imaginary part of density-density response function $\chi_{nn}(k,\omega)$ to the instantaneous structure factor $S(k)$. At zero temperature the FDT reads~\cite{Giuliani_and_Vignale}
\be\label{eq:fluc}
S(k)= -\frac{\hbar}{\pi n}\int_0^\infty d\omega~ \Im m\left[ \chi_{nn}(k,\omega)\right]~.
\ee

To make some progress, we neglect the frequency dependence of the effective potential $\psi(k,\omega)$ in Eq.~(\ref{eq:chinn}): we replace the complex function $\psi(k,\omega)$ by a real quantity, which we denote by the symbol ${\bar \psi}(k)$. This approximation is often made in treating correlation effects in the electron liquid~\cite{Pines_and_Nozieres,Giuliani_and_Vignale} and is certainly shared by the most elementary theories based on LFFs (Hubbard and STLS). In this case, one can view Eq.~(\ref{eq:fluc}) as an integral equation for the unknown quantity ${\bar \psi}(k)$, assuming that the l.h.s. of Eq.~(\ref{eq:fluc}), {\it i.e.} the static structure factor, is accurately known, {\it e.g.} from QMC simulations or microscopic theories such as the one outlined in Sect.~\ref{sect:theory}. This fully numerical approach has been successfully employed in different contexts~\cite{boronat_prl_2003,asgari_prb_2006}. The physical interpretation of ${\bar \psi}(k)$ is clear: it represents the ``best" average effective potential [averaged over frequency, as from Eq.~(\ref{eq:fluc})] which, by virtue of the FDT, makes the response of the system {\it consistent} with the local structure of the fluid around a reference dipole (the Pauli-Coulomb hole).

In the spirit of making the problem at hand more amenable to a semi-analytical treatment, we also use the so-called ``mean-spherical
approximation" (MSA) for $\chi_0(k,\omega)$~\cite{asgari_prb_2006}:
\be\label{eq:chi_MSA}
\chi^{({\rm MSA})}_0(k,\omega) \equiv \frac{2 n \epsilon(k)}{(\hbar\omega+i 0^+)^2-\left[\epsilon(k)/S_{\rm HF}(k)\right]^2}~,
\ee
where $S_{\rm HF}(k)$ is the well-known 2D HF static structure factor~\cite{Giuliani_and_Vignale}.
This approximation allows us to perform the integration over $\omega$ in
Eq.~(\ref{eq:fluc}) analytically, yielding
\be\label{eq:veff_MSA}
{\bar \psi}(k) \stackrel{\rm MSA}{=} \frac{\epsilon(k)}{2 n}\left[ \frac{1}{S^2(k)}- \frac{1}{S^2_{\rm HF}(k)} \right]~.
\ee
For the static structure factor $S(k)$ in the r.h.s. of Eq.~(\ref{eq:veff_MSA}) we use the FHNC theory described above in Sect.~\ref{sect:theory}.

We can now use Eqs.~(\ref{eq:chinn}) and~(\ref{eq:veff_MSA}) to address two important issues.

First, we can carry out  a linear-stability analysis of the liquid phase against density modulations. In this respect, a pole in the {\it static} density-density response function $\chi_{nn}(k_{\rm c},\omega=0)$ at a finite wave vector $k_{\rm c}$ signals an instability of the liquid state towards a density wave with period $\propto k^{-1}_{\rm c}$. In practice, we need to find whether the following equation,
\be\label{eq:CDW}
\varepsilon(k,\omega=0) = 1- {\bar \psi}(k)\chi_0(k,\omega=0)=0~,
\ee
admits a solution at a finite wave vector $k_{\rm c}$. We remind the reader that, in the static limit, $\chi_0(k,\omega)$ is purely real.

Second, we can study the existence of a collective mode~\cite{Giuliani_and_Vignale} in the density channel (zero sound~\cite{Pines_and_Nozieres}). This is the solution of the complex equation $\varepsilon(k,\omega) =0$ or, equivalently, of the following two real equations:
\be\label{eq:mode}
\left\{
\begin{array}{l}
1- {\bar \psi}(k)\Re e\left[\chi_0(k,\omega)\right]=0\vspace{0.2 cm}\\
\Im m\left[\chi_0(k,\omega)\right]=0
\end{array}
\right.~.
\ee
The solution $\omega_{\rm ZS} = \omega_{\rm ZS}(k)$ of Eq.~(\ref{eq:mode}) corresponds to a self-sustained oscillation with a non-trivial dispersion relation and a finite velocity $v_{\rm ZS} \equiv \lim_{k\to 0} \omega_{\rm ZS}(k)/k$ in the long-wavelength limit.  The second equation means that the collective mode is undamped when it falls in the region of $(k,\omega)$ space where particle-hole pairs are absent. This occurs when $v_{\rm ZS}> v_{\rm F}$, $v_{\rm F} = \hbar k_{\rm F}/m$ being the Fermi velocity. When the collective mode enters the particle-hole continuum, Landau damping starts: the mode has sufficient energy to decay by emitting a particle-hole pair while, at the same time, conserving momentum.

Before concluding this Section, we derive a formal expression for the ZS velocity, $v_{\rm ZS}$, in terms of ${\bar \psi}(0)$. (As we will see below in Sect.~\ref{sect:results}, ${\bar \psi}(k)$ is regular and positive at $k=0$.) In order to find $v_{\rm ZS}$ we use the following long-wavelength limit
of the ideal response function~\cite{santoro_prb_1988,Giuliani_and_Vignale}:
\be\label{eq:asymptoticlimit}
\lim_{k\to 0} \Re e~[\chi_0(k, v_{\rm F} k \nu)] = -N(0)\left(1-\frac{\nu}{\sqrt{\nu^2-1}}\right)~,
\ee
where $N(0) = m/(2\pi \hbar^2)$ is the 2D density-of-states at the Fermi energy and $\nu = \omega/(v_{\rm F} k) = {\rm constant}$. Note that, in Eq.~(\ref{eq:asymptoticlimit}), the ratio between $\omega = v_{\rm F} k \nu$ and $k$ remains {\it constant} in the limit $k\to 0$, precisely as in the ZS mode [$\lim_{k\to 0} \omega_{\rm ZS}(k)/k = {\rm constant}$]. It is very important to observe that the asymptotic behavior (\ref{eq:asymptoticlimit}) needed for the calculation of the ZS velocity is very different from the usual high-frequency limit imposed by the f-sum rule~\cite{Giuliani_and_Vignale}:
\be\label{eq:fsumrule}
\Re e~[\chi_0(k, \omega \gg v_{\rm F} k)] = \frac{n k^2}{m \omega^2}~.
\ee

Now, replacing Eq.~(\ref{eq:asymptoticlimit}) [and {\it not} Eq.~(\ref{eq:fsumrule})] in Eq.~(\ref{eq:mode}), we find the following formal expression for the ZS velocity in units of the Fermi velocity:
\be\label{eq:soundvelfinal}
\frac{v_{\rm ZS}}{v_{\rm F}} = \frac{1+N(0){\bar \psi}(0)}{\displaystyle \sqrt{1+2 N(0){\bar \psi}(0)}}~,
\ee
which is well defined if ${\bar \psi}(0) > -[2N(0)]^{-1}$. Note that the quantity on the r.h.s. of Eq.~(\ref{eq:soundvelfinal}) is always larger than one. We therefore conclude that, within the approximations we made to derive Eq.~(\ref{eq:veff_MSA}), a 2D DFG displays {\it always} ({\it i.e.} for every value of the coupling constant $\lambda$) an undamped ZS mode, in agreement with Ref.~\cite{lu_pra_2012}. As discussed at length in Ref.~\cite{lu_pra_2012}, this mode stems entirely from correlation effects and it is thus not describable within the HF approximation. However, the RPA, which is the minimal theory including correlations, is not enough in this respect since it yields a ZS mode with a velocity that depends on the short-range cut off of the bare dipole-dipole interaction~\cite{li_prb_2010}. A serious theory of the ZS mode thus requires the inclusion of correlation effects {\it beyond} RPA. The FHNC theory discussed in this Article is an example.

\section{Numerical results and discussion}
\label{sect:results}
In this Section we present our main numerical results.

\begin{figure}
\includegraphics[width=1.0\linewidth]{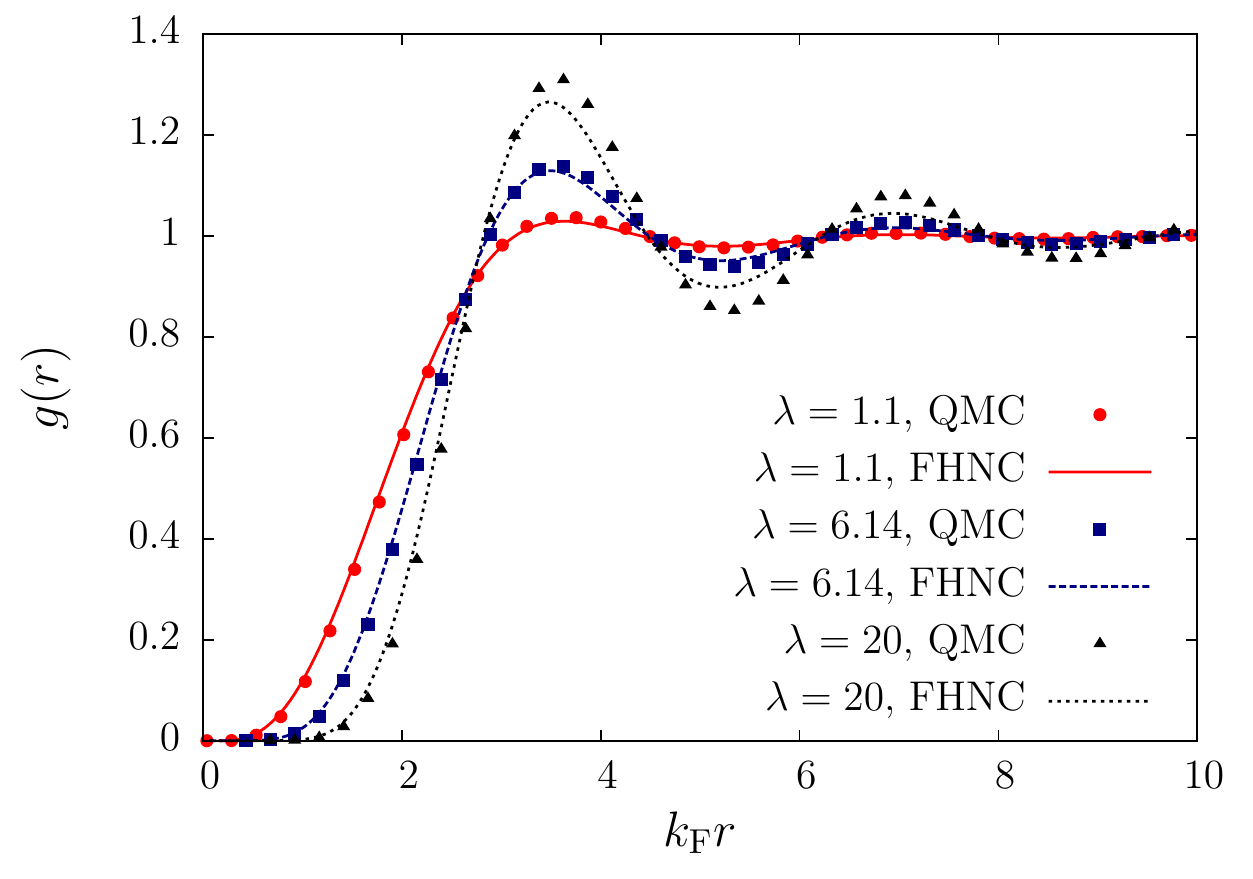}
\caption{(Color online) The pair distribution function $g(r)$ of a 2D fluid of dipolar fermions is plotted
as a function of the scaled distance $k_{\rm F} r$ and for various values of the dimensionless coupling constant $\lambda$. In this plot, lines label the results of the FHNC approximation (this work) while symbols label QMC results~\cite{matveeva_arxiv_2012}.\label{fig:gr}}
\end{figure}

\begin{figure}
\includegraphics[width=1.0\linewidth]{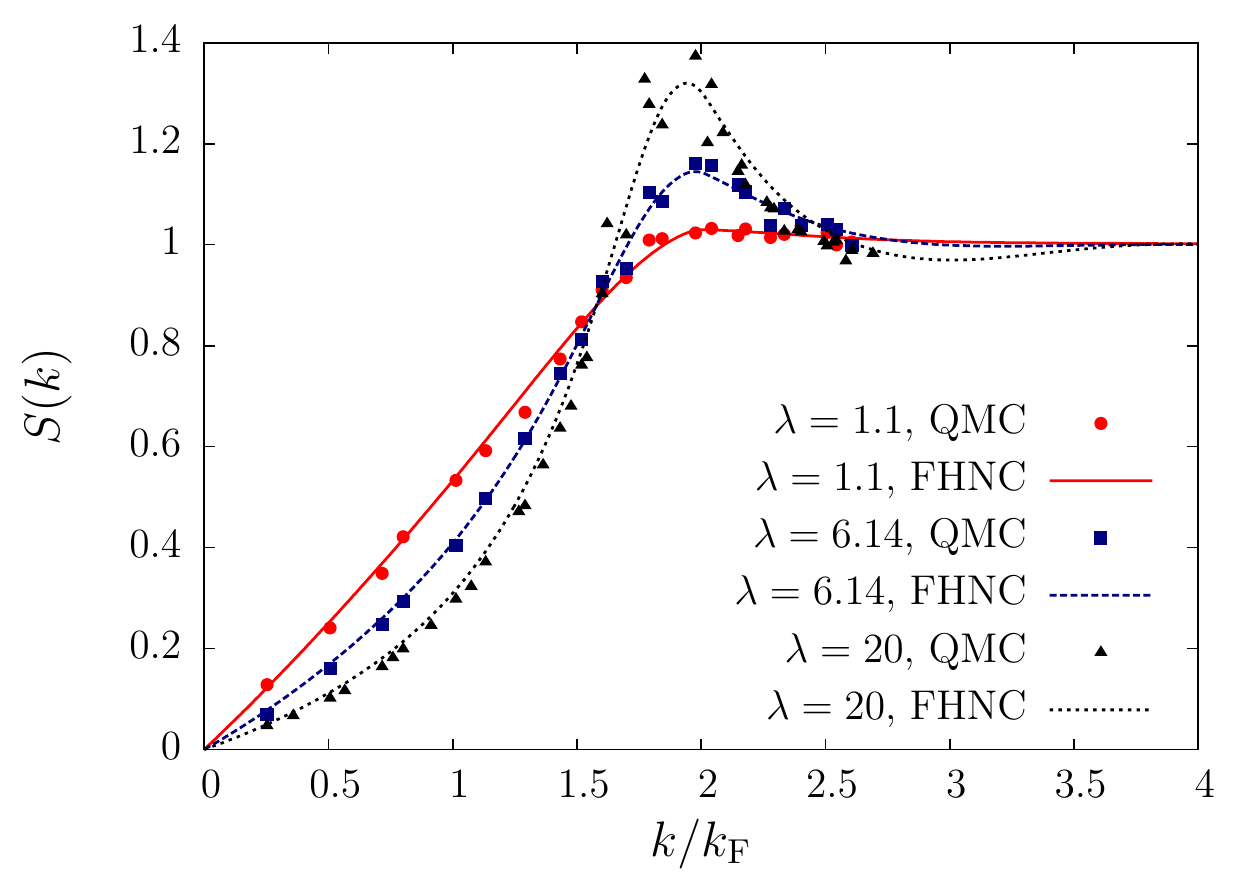}
\caption{(Color online) The instantaneous structure factor $S(k)$ of a 2D fluid of dipolar fermions is plotted as a function of $k/k_{\rm F}$ and
for various values of $\lambda$. In this plot, lines label the results of the FHNC approximation (this work) while
symbols label QMC results~\cite{matveeva_arxiv_2012}.\label{fig:sk}}
\end{figure}

We begin by showing our results for the PDF $g(r)$ and static structure factor $S(k)$.
In Figs.~\ref{fig:gr} and~\ref{fig:sk} we compare our results (lines) with the corresponding QMC data (symbols)~\cite{matveeva_arxiv_2012}.
The agreement between theory and numerical simulations is clearly excellent up to very large values of the dimensionless coupling constant
$\lambda$ ($\lambda = 20$). At these values of $\lambda$, conventional theories such as RPA and STLS fail even qualitatively. Note that, according to the QMC study by Matveeva and Giorgini~\cite{matveeva_arxiv_2012}, a liquid-to-crystal quantum phase transition is expected to occur at $\lambda \sim 25$. This is clearly signaled by the amplitude of the first-neighbor peak in the static structure factor (see Fig.~\ref{fig:sk}), which increases with increasing $\lambda$ indicating the build up of correlations in the liquid phase upon approaching crystalline order.
\begin{figure}
\includegraphics[width=1.0\linewidth]{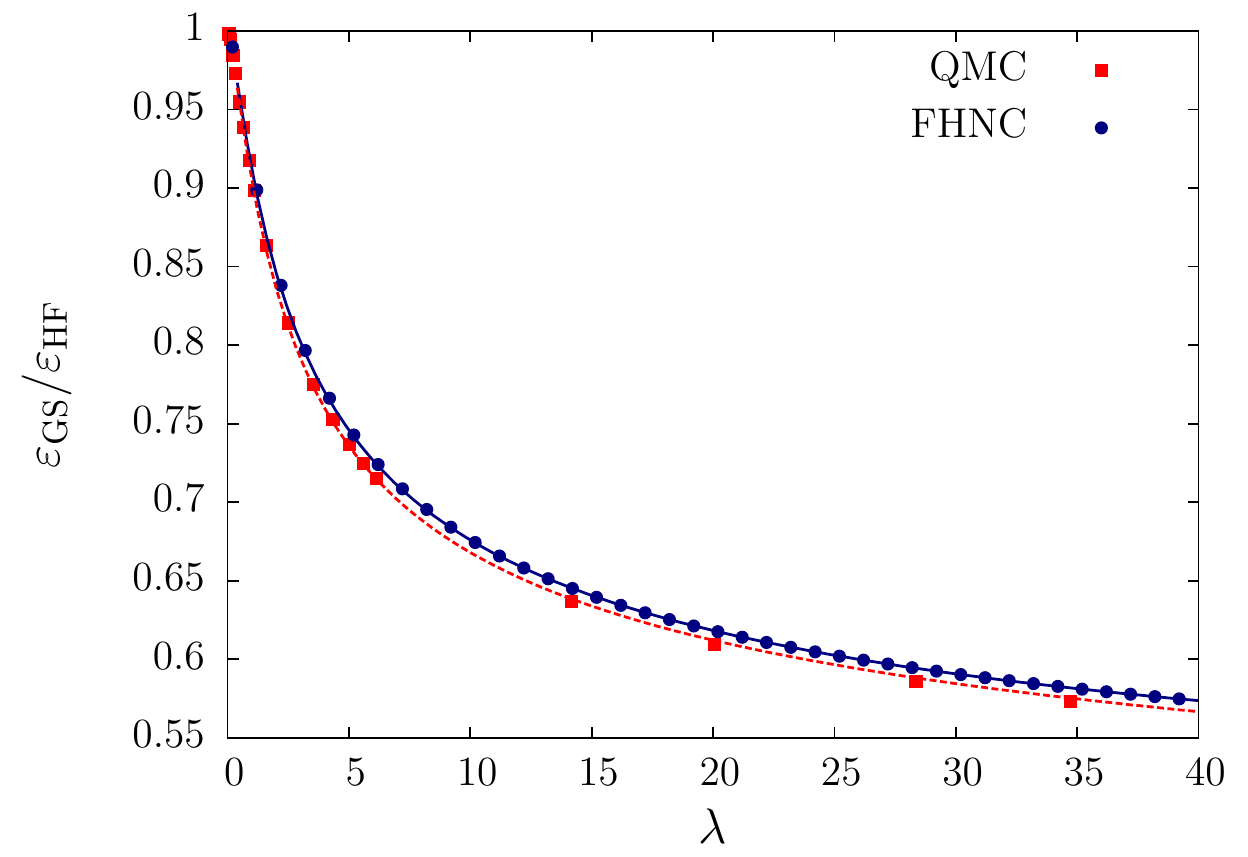}
\caption{(Color online) The ground-state energy of a 2D fluid of dipolar fermions [in units of the Hartree-Fock energy $\varepsilon_{\rm HF}/\varepsilon_0 =  1+128 \lambda/(45\pi)$], is plotted as a function of $\lambda$. Blue circles label the results of the FHNC approximation (this work) while red squares label the QMC results~\cite{matveeva_arxiv_2012}. The solid line represents the parametrization formula in Eq.~(\ref{eq:e_gs_fit}) with $a = 1.5006$, $b=1.0107$, and $c=0$. The dashed line represents the parametrization in Eq.~(\ref{eq:e_gs_fit}) with $a = \sqrt{\zeta_3}$ (see main text), $b=1.1017$, and $c=-0.0100$.\label{fig:energy}}
\end{figure}

The PDF shown in Fig.~\ref{fig:gr} can be used to calculate the ground-state energy by employing Eq.~(\ref{eq:e_gs}).
In Fig.~\ref{fig:energy} we report our results for the ground-state energy as obtained
from the PDF calculated at the FHNC level. In the same plot we compare our findings with the corresponding QMC results~\cite{matveeva_arxiv_2012}. In passing, we note that our FHNC results for the ground-state energy (per particle) can be accurately parametrized by the following expression:
\ber\label{eq:e_gs_fit}
\varepsilon_{\rm GS}(\lambda) &=&\varepsilon_{\rm 0}
\Bigg[1+\frac{128}{45\pi}\lambda \nonumber\\
&-& \frac{\lambda^2}{2}\ln{\left(1+\frac{1}{a~\sqrt{\lambda}+b~\lambda +c~\lambda^{3/2}}\right)}
\Bigg]~,
\eer
where $a$, $b$, and $c$ are numerical constants. The sum of the first two terms in square brackets on the r.h.s. of Eq.~(\ref{eq:e_gs_fit}) yields the HF approximation for the ground-state energy~\cite{lu_pra_2012}: $\varepsilon_{\rm HF} \equiv \varepsilon_0[1+128 \lambda/(45\pi)]$. The best fit of our FHNC data for the energy of the liquid phase up to $\lambda = 40$ is obtained by using $a$ and $b$ as free fitting parameters and setting $c=0$:
we find $a=1.5006$ and $b=1.0107$. The result of this two-parameter fit is shown in Fig.~\ref{fig:energy} (solid line).

Alternatively, the simple formula in Eq.~(\ref{eq:e_gs_fit}) can be used to parametrize also the QMC data by Matveeva and Giorgini~\cite{matveeva_arxiv_2012}.
Since these data are believed to be essentially exact, we can fix the value of $a$ by imposing that Eq.~(\ref{eq:e_gs_fit}) reproduces exactly the results of second-order perturbation theory~\cite{lu_pra_2012}. Straightforward algebraic manipulations on Eq.~(\ref{eq:e_gs_fit}) yield the following expansion in powers of $\lambda$ for $\lambda \to 0$:
\be\label{eq:expansion}
\varepsilon_{\rm GS}(\lambda) =\varepsilon_{\rm 0}\left[1+\frac{128}{45\pi}\lambda + \frac{\lambda^2}{4}\ln(a^2\lambda) + \dots\right]~,
\ee
where ``$\dots$" denotes higher-order terms. To the same order of perturbation theory, Lu and Shlyapnikov~\cite{lu_pra_2012} find [Eq.~(91) in their work]:
\be\label{eq:expansionexact}
\varepsilon_{\rm GS}(\lambda) =\varepsilon_{\rm 0}\left[1+\frac{128}{45\pi}\lambda + \frac{\lambda^2}{4}\ln(\zeta_3\lambda) + \dots\right]
\ee
where $\zeta_3 = 1.43$ (we have taken the limit $A \to 0$ in the expression for $\zeta_3$ given in Ref.~\cite{lu_pra_2012}).
Comparing Eq.~(\ref{eq:expansion}) with Eq.~(\ref{eq:expansionexact}) we conclude that $a = \sqrt{\zeta_3} \sim 1.2$. The parameters $b$ and $c$
can then be used to yield the best fit to the QMC data for the energy of the liquid phase up to $\lambda =72$~\cite{matveeva_arxiv_2012}: we find $b =1.1017$ and $c=-0.0100$.
The result of this two-parameter fit is also shown in Fig.~\ref{fig:energy} (dashed line).

 The difference between the total ground-state energy and the non-interacting contribution $\varepsilon_0$ defines the interaction energy: $\varepsilon_{\rm int}(\lambda) = \varepsilon_{\rm GS}(\lambda) - \varepsilon_{\rm 0}(\lambda)$. 
Note that unlike the gellium model for electron gases~\cite{Giuliani_and_Vignale}, the Hartree contribution to the interaction energy does not vanish in our system of polarized DFGs~\cite{lu_pra_2012}.
Eq.~(\ref{eq:e_gs_fit}) thus provides an extremely useful input for calculations
of ground-state properties of inhomogenous 2D DFGs based on density functional theory (DFT)~\cite{Giuliani_and_Vignale}.
In DFT, indeed, one needs to approximate the unknown interaction energy $E_{\rm int}[n({\bm r})]$, viewed as a functional of the local ground-state density $n({\bm r})$.
In the local density approximation (LDA) one can write~\cite{Giuliani_and_Vignale}
\be\label{eq:LDA}
E_{\rm int}[n({\bm r})] \stackrel{\rm LDA}{=} \int d^2{\bm r}~n({\bm r})\varepsilon_{\rm int}(\lambda({\bm r}))~,
\ee
where $\lambda({\bm r})$ is defined as in Eq.~(\ref{eq:couplingconstant}) with $n$ replaced by the local density $n({\bm r})$. An example where the DFT-LDA approach could be very useful is a 2D DFG in the presence of an in-plane harmonic confinement potential $V_{\rm ext} = \sum_i m\omega^2 {\bm r}^2_i/2$.

\begin{figure}
\includegraphics[width=1.0\linewidth]{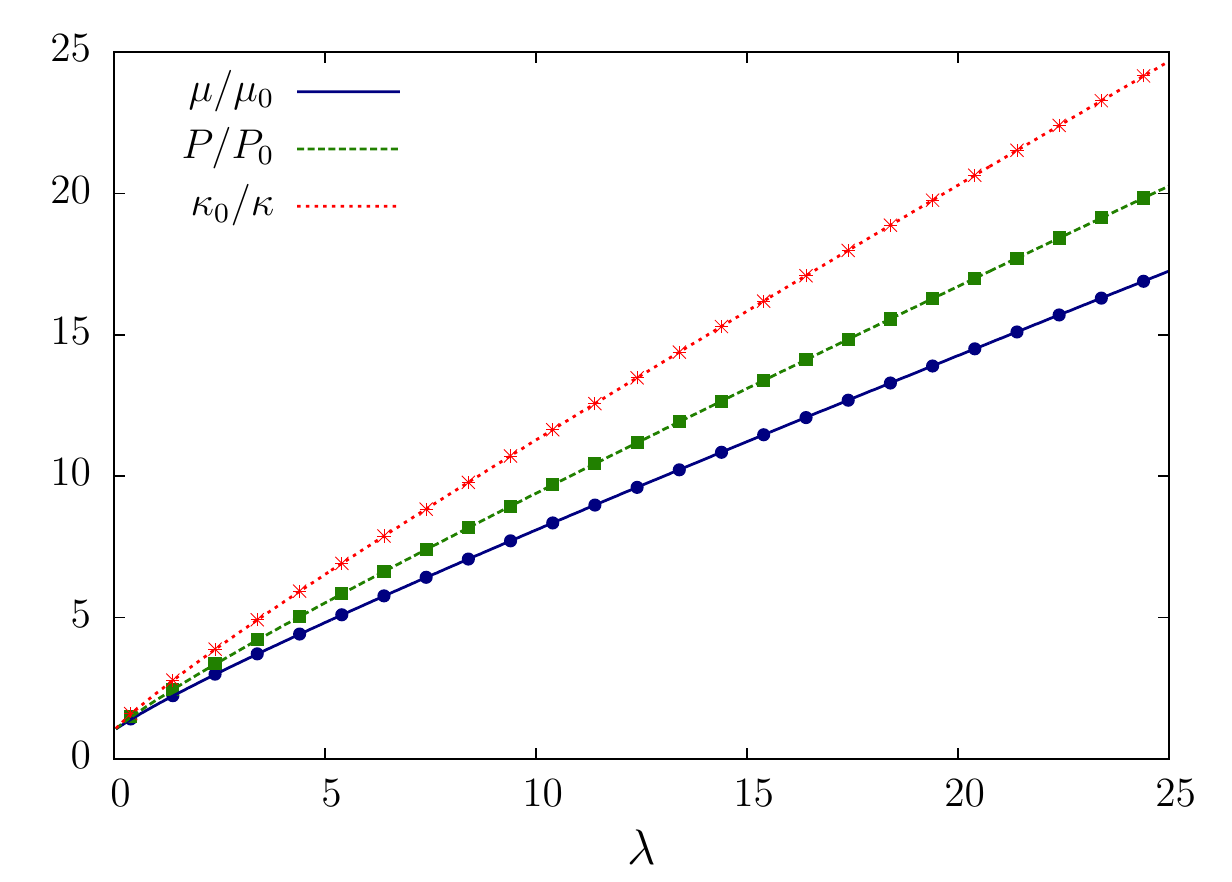}
\caption{(Color online) Three important thermodynamic quantities: the chemical potential $\mu$,  pressure $P$, and inverse compressibility $\kappa^{-1}$ of a 2D fluid of dipolar fermions (in units of their non-interacting values) are plotted as functions
of $\lambda$. Lines label analytic results obtained from the parametrization formula (\ref{eq:e_gs_fit}) while symbols label numerical results obtained directly from the FHNC ground-state energy.\label{fig:thermo}}
\end{figure}

From the knowledge of the ground-state energy (per particle)
$\varepsilon_{\rm GS}$ we can also construct a number of thermodynamic
quantities at zero temperature. Most notably, the chemical
potential $\mu=\partial(n\varepsilon_{\rm GS})/\partial n$,
the pressure $P=n^2\partial\varepsilon_{\rm GS}/\partial n$, and the
inverse compressibility $\kappa^{-1}=n\partial P/\partial n$
are readily obtained from the interpolation formula given in Eq.~(\ref{eq:e_gs_fit}).
We display these quantities as functions of the interaction strength $\lambda$ in
Fig.~\ref{fig:thermo}. Note that all these quantities, which still remain to be experimentally measured,
are strongly {\it enhanced} by interactions.

\begin{figure}
\includegraphics[width=1.0\linewidth]{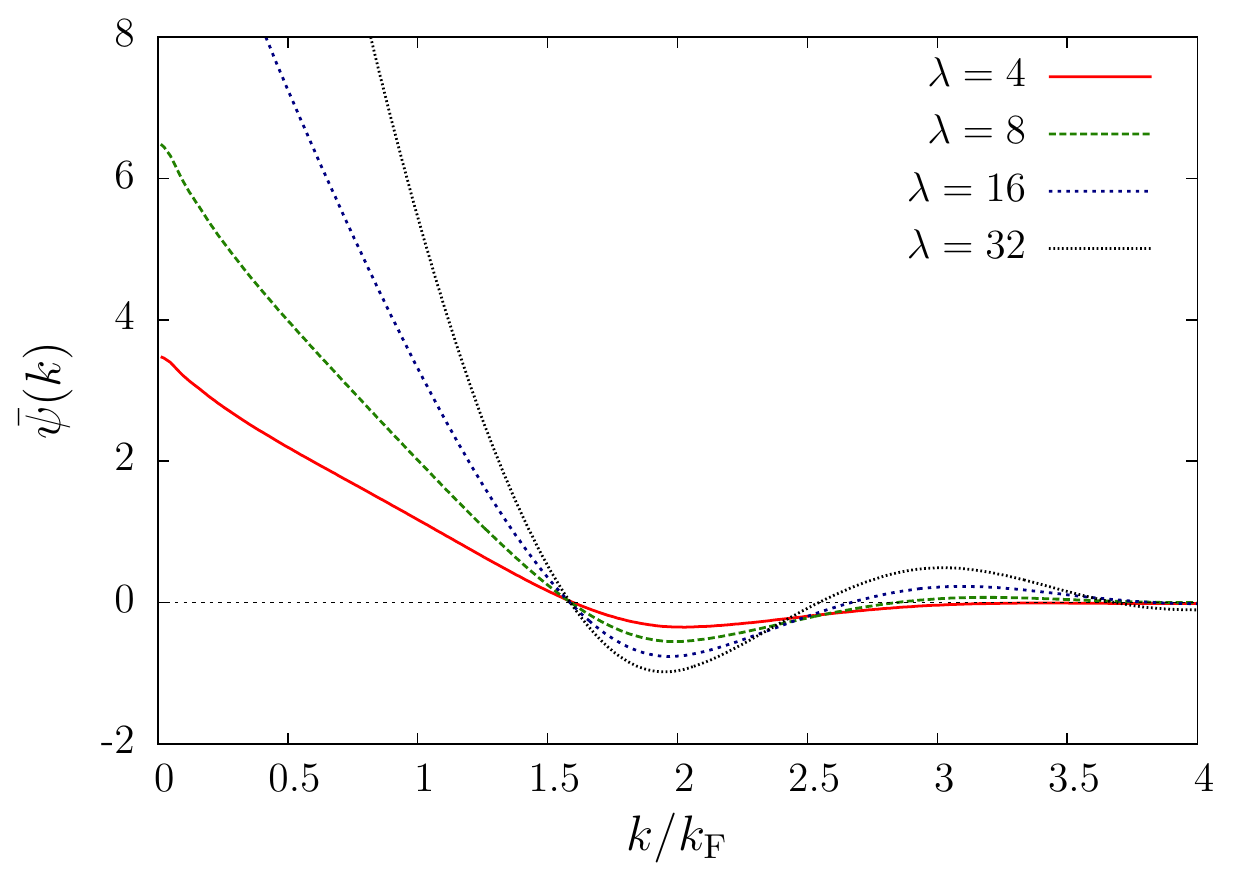}
\caption{(Color online) The effective interaction ${\bar \psi}(k)$ [in units of $2\pi\hbar^2/m$]
in a 2D fluid of dipolar fermions as obtained from Eq.~(\ref{eq:veff_MSA}) is plotted as a function of $k/k_{\rm F}$
for various values of $\lambda$. Note that ${\bar \psi}(k\to 0)$ tends to a positive value.
\label{fig:veff}}
\end{figure}

Fig.~\ref{fig:veff} illustrates the effective potential ${\bar \psi}(k)$ as obtained from Eq.~(\ref{eq:veff_MSA}).
We clearly see from this plot that ${\bar \psi}(k)$ is regular and positive at $k =0$.

\begin{figure}
\includegraphics[width=1.0\linewidth]{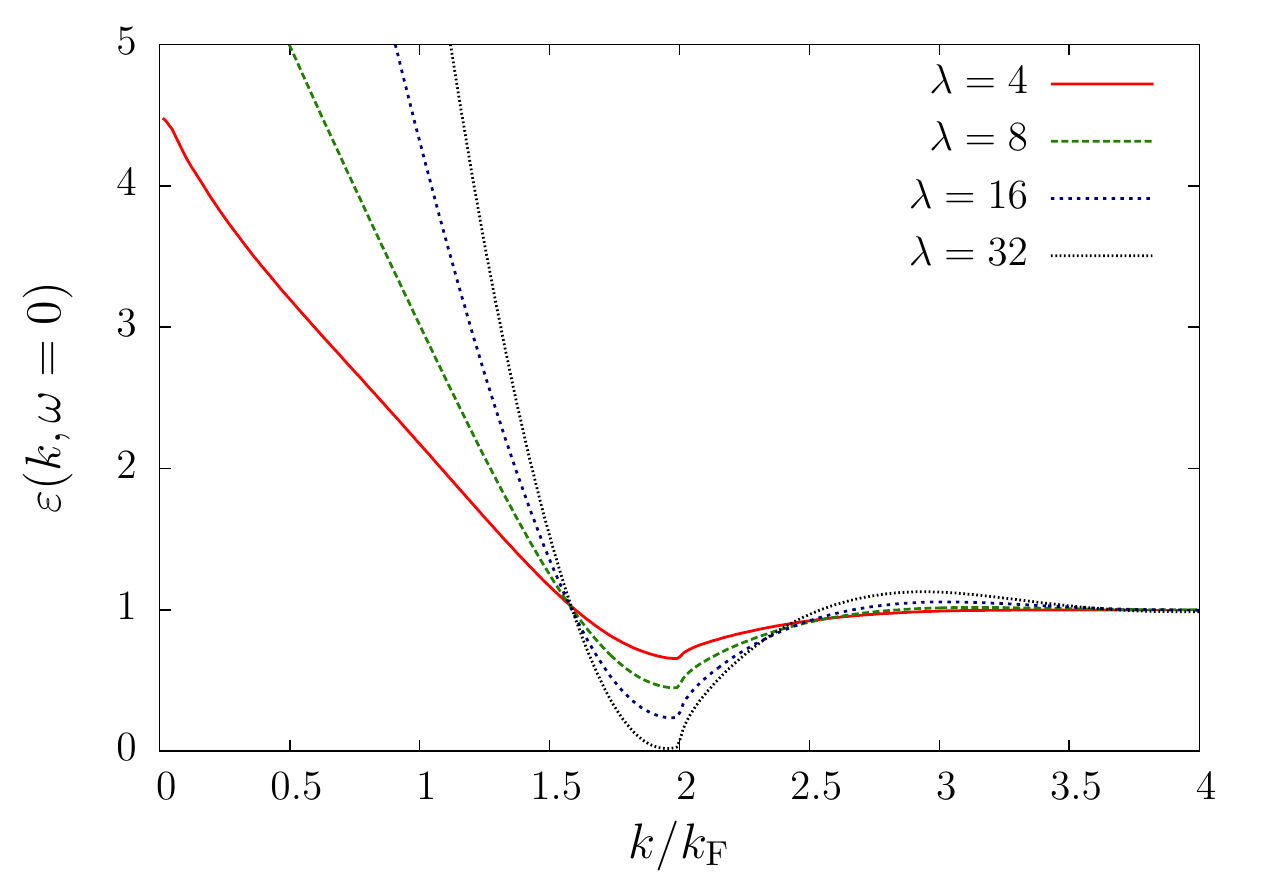}
\caption{(Color online) The static dielectric function $\varepsilon(k, \omega=0)$
of a 2D fluid of dipolar fermions as a function of $k/k_{\rm F}$ and for various values of $\lambda$.
Note that $\varepsilon(k, \omega=0)$ never vanishes, even for very large values of $\lambda$.
\label{fig:epsilon}}
\end{figure}

In Fig.~\ref{fig:epsilon} we plot $\varepsilon(k,0) = 1- {\bar \psi}(k)\chi_0(k,\omega=0)$
as a function of wave vector $k$ and for different values of $\lambda$. Increasing the interaction strength, a
minimum occurs in $\varepsilon(k,0)$ (yielding a peak in the density-density response function)
at a wave vector close to $2 k_{\rm F}$. This minimum remains finite, though, up to the largest value of $\lambda$ we have investigated ($\lambda = 40$). In other words, our theory does not predict any density-wave instability in a 2D DFG.
This is in agreement with the QMC results by Matveeva and Giorgini~\cite{matveeva_arxiv_2012},
who have shown that a stripe phase has higher energy than that of liquid and crystal phases at any $\lambda$.

\begin{figure}
\includegraphics[width=1.0\linewidth]{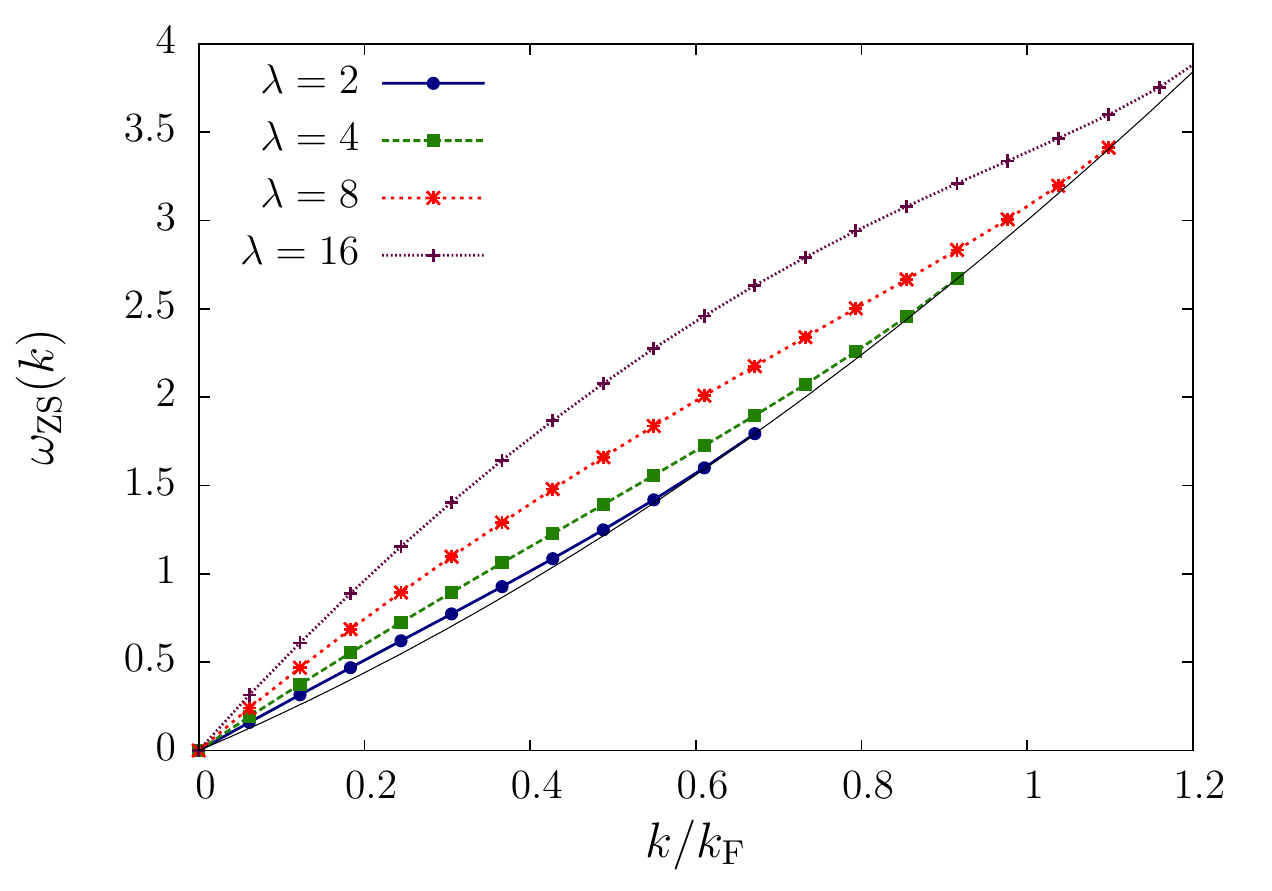}
\caption{(Color online) The frequency $\omega_{\rm ZS}(k)$ of the zero-sound mode in a 2D fluid of dipolar fermions is plotted as a function of $k/k_{\rm F}$ for various values of $\lambda$. The thin (black) solid line represents the upper bound of the particle-hole (p-h) continuum, {\it i.e.} $\omega_+(k) = \epsilon(k)/\hbar + v_{\rm F} k$. Note that the zero-sound mode lies above the p-h continuum for every $\lambda$, up to a $\lambda$-dependent critical wave vector at which Landau damping starts.\label{fig:ZS}}
\end{figure}

Finally, in Fig.~\ref{fig:ZS} we illustrate our predictions for the dispersion of the ZS mode.
As already discussed at the end of Sect.~\ref{sect:CDW}, our theory predicts an undamped ZS mode at long wavelengths for every value of $\lambda$. The ZS velocity as well as the critical wave vector at which Landau damping starts increase with increasing $\lambda$.
\section{Summary}\label{sect:concl}

In summary, we have presented an extensive study of ground-state and dynamical properties of a strongly correlated two-dimensional spin-polarized fluid of dipolar fermions.

The main focus of our work has been on the pair distribution function $g(r)$, a key ground-state property of any quantum fluid. To calculate the pair distribution function
we have employed the Fermi-hypernetted-chain approximation combined with a zero-energy scattering Schr\"odinger equation for the
``pair amplitude" $\sqrt{g(r)}$. The effective potential that enters this equation includes a bosonic term from Jastrow-Feenberg correlations
and a fermionic contribution from kinetic energy and exchange, which is tailored
to reproduce the Hartree-Fock limit at weak coupling. Our results for the pair distribution function and the static structure factor $S(k)$ have been severely benchmarked against state-of-the-art quantum Monte Carlo results by Matveeva and Giorgini~\cite{matveeva_arxiv_2012}. Very good agreement with these results has been achieved over a wide range of coupling constants up to the liquid-to-crystal quantum phase transition.

By combining our knowledge on the pair distribution function with the fluctuation-dissipation theorem, we have been able to calculate in an approximate fashion also
the dynamical density-density response function. This ingredient has been used to demonstrate that, in a two-dimensional spin-polarized fluid of dipolar fermions, i) the liquid phase is stable towards the formation of density waves up to the liquid-to-crystal quantum phase transition (in agreement with Ref.~\cite{matveeva_arxiv_2012}) and ii) an undamped zero-sound mode occurs for any value of the interaction strength, down to infinitesimally weak couplings (in agreement with Ref.~\cite{lu_pra_2012}).

Last but not least, we have presented a useful parametrization formula, Eq.~(\ref{eq:e_gs_fit}), for the ground-state energy of a two-dimensional spin-polarized fluid of dipolar fermions, which fits well both our Fermi-hypernetted-chain results and the quantum Monte Carlo data by Matveeva and Giorgini~\cite{matveeva_arxiv_2012}. This can be very effectively employed in density-functional calculations of 2D
inhomogenous dipolar fermions.

\acknowledgments
We are indebted to Natalia Matveeva and Stefano Giorgini for providing
us with their QMC data. It is also a pleasure to thank Nikolaj Zinner for useful discussions.
S.H.A. gratefully acknowledges the kind hospitality of the IPM in
Tehran, Iran during the final stages of this work. B.T.
acknowledges support from TUBITAK (through grants 109T267, 209T059,
210T050) and TUBA.

\end{document}